\documentclass[a4paper,12pt]{article}
\usepackage{amsmath,amsfonts,amssymb,bm,amsthm}
\usepackage{mathrsfs}
\usepackage{cases}
\usepackage{cite}
\usepackage{indentfirst}
\usepackage[usenames]{color}
\usepackage{hyperref}
\usepackage{fancyhdr}
\usepackage{geometry}
\definecolor{webgreen}{rgb}{0,.5,0}
\definecolor{webbrown}{rgb}{.6,0,0}

\usepackage{color}

\newtheorem{theorem}{Theorem}

\newtheorem{remark}{Remark}

\newtheorem{definition}{Definition}

\newtheorem{corollary}{Corollary}

\allowdisplaybreaks

\begin{document}

\title{The generalized covariant Hamilton system in complex coordinates}

\author{ Gen  WANG \thanks{HS: Hamilton System; \newline GCHS: Generalized Covariant Hamilton System; \newline GSPB:Generalized structural Poisson bracket; \newline  TGHS: Thorough generalized Hamiltonian System} }

\date{\em{\small{ School of Mathematical Sciences, Xiamen University,\\
     Xiamen, 361005, P.R.China}} }

\maketitle

\begin{abstract}
Imitating methods of working on the GCHS and GSPB defined on ${{\mathbb{R}}^{r}}$ in real coordinates, an attempt to follow this way in complex coordinates is considered, then we try to generalize the PB on ${{\mathbb{C}}^{n}}$ in complex coordinates to the GSPB with zero restriction of the non-degeneracy expressed in complex coordinates that is compatible with the real case in formulas. Thusly, then the GCHS in complex coordinates defined by the GSPB is self-consistent to the real situation. Meanwhile, we find some difference of the GCHS between the real and complex case in formula. Much of what distinguishes a GCHS in real coordinates from a GCHS in complex coordinates is that the TGHS in complex form has an extra expression.

\end{abstract}

\newpage
\tableofcontents

\section{Introduction}
Classical Hamilton system (HS or CHS) \footnote{HS: Hamilton System; PB:Poisson bracket\\GPB:Generalized Poisson bracket; GHS:Generalized Hamilton System}  theory is defined on the even-dimensional phase space ${{\mathbb{R}}^{2n}}$ which is corresponding to the classical Poisson bracket (PB or CPB). Although this structure has good properties, but it also restricts its application in science fields. Therefore, scientists \cite{1,2} still have been trying to generalize the classical Hamilton mechanics theory in order to widen its applications for completely explaining more problems. Such generalized theory of Hamilton system has been called generalized Hamilton system (GHS). Scientists have used the generalized Poisson bracket (GPB) to define generalized Hamilton system (GHS) \cite{3,4} on ${{\mathbb{R}}^{r}}$ which means arbitrary dimensional,  which is more concise and convenient,  thus then it's promoting the further development of Hamilton mechanics. The phase space of the generalized Hamilton system can be any finite dimensional or even infinite dimensional. The phase space of the generalized Hamilton system is a Poisson manifold. In other aspects, the HS expressed in complex coordinates is also an important subject in mathematics and physics, and it has developed well.  Recently, \cite{5} has proposed the generalized structural Poisson bracket (GSPB) for the generalized covariant Hamilton system (GCHS) on ${{\mathbb{R}}^{r}}$ as a complete theory to solve some left questions of GHS. Similarly, the phase space of the generalized covariant Hamilton system also can be any finite dimensional or even infinite dimensional. The phase space of the generalized covariant Hamilton system is a generalized Poisson manifold equipped with generalized structural Poisson bracket (GSPB).

The aim of this paper is to prove that GCHS defined by GSPB in complex coordinates which contains dynamic subsystem: TGHS and S-dynamics is compatible with the real coordinates.  Confidently and certainly, this is a self-consistent and complete perfect system. We discover that the biggest difference between the GCHS in real coordinates and its complex coordinates form is manifested by the TGHS. More importantly, the framework of GCHS can surely answer the prime question in GHS,  the conditions of GHS are not enough to picture a nonlinear system in some levels,  but it can be expressed as a GCHS for Hamiltonian realization and nonlinear system.

\subsection{Poisson bracket in real coordinates}
The research on Hamilton's equations has occurred along two directions. One direction is concerned with the geometrical structure of Hamilton's equations. The other direction is linked to the dynamical properties of the flow generated by Hamiltonian vector fields.

For a $C^{r}$, $r \geq 2$, real valued function on some open set $U \subset \mathbb{R}^{2 n}$,  Hamilton's canonical equations are given by \cite{3} $$\dot{q}=\frac{\partial H}{\partial p}(q, p),~~ \dot{p}=-\frac{\partial H}{\partial q}(q, p), \quad(q, p) \in U \subset \mathbb{R}^{2 n}$$
It will often be convenient to write these in a more compact notation. Defining $x \equiv(q, p)$, then above equations can be written as
$$\dot{x}=J\nabla H(x)$$
where $J\left( x \right)$ is structure matrix, $\nabla H\left( x \right)$  is the gradient of the function Hamilton.  More specifically,
$$J=\left(\begin{array}{cc}{0} & {\text { id }} \\ {-\text { id }} & {0}\end{array}\right)$$where
 "id" denotes the $n \times n$ identity matrix, and $\nabla H(x) \equiv\left(\frac{\partial H}{\partial q}, \frac{\partial H}{\partial p}\right)$.  General notation for Hamiltonian vector fields, for the sake of a more compact notation,  we will occasionally denote the Hamiltonian vector field derived from a function $H$ by
$$X_{H}(x) \equiv\left(\frac{\partial H}{\partial p},-\frac{\partial H}{\partial q}\right)$$
Let $H, G: U \rightarrow \mathbb{R}$ denote two  $C^{r}, r \geq 2$, functions. Then the Poisson bracket of these two functions is another function, and it is defined through the symplectic form as follows: \cite{3}
\begin{equation}\label{eq18}
  \{H, f\}_{PB} \equiv \Omega\left(X_{H}, X_{f}\right) \equiv\left\langle X_{H}, J X_{f}\right\rangle
\end{equation}
It follows immediately that the Poisson bracket is antisymmetric. Using the  definitions of  $X_{H}, X_{f}$,  as well as the canonical symplectic form, we easily see that above symplectic form assumes the following coordinate form:
\begin{equation}\label{eq3}
  \left\{ H,f\right\}_{PB}=\sum\limits_{j}{\left( \frac{\partial H}{\partial {{q}^{j}}}\frac{\partial f}{\partial {{p}^{j}}}-\frac{\partial H}{\partial {{p}^{j}}}\frac{\partial f}{\partial {{q}^{j}}} \right)}
\end{equation}
Then the Poisson bracket of these two functions is another function. The classical Poisson bracket defined on functions on ${{\mathbb{R}}^{2n}}$ is \cite{3,4,5}
 \[\left\{ {{f}_{1}},{{f}_{2}} \right\}_{PB}=\sum\limits_{i}{\left(  \frac{\partial {{f}_{1}}}{\partial {{q}^{i}}}\frac{\partial {{f}_{2}}}{\partial {{p}^{i}}}-\frac{\partial {{f}_{1}}}{\partial {{p}^{i}}}\frac{\partial {{f}_{2}}}{\partial {{q}^{i}}} \right)},~~~\forall {{f}_{j}}\in {{C}^{\infty }}\left( M\right)\]
In finite dimensions, Hamilton's equations in canonical coordinates are
\[{\dot {q}}^{i}=\frac{\partial H}{\partial {{p}^{i}}},~~{\dot {p}}^{i}=-\frac{\partial H}{\partial {{q}^{i}}},~~i=1,\cdots ,n\]
By using PB Eq \eqref{eq3}, it follows
\[{\dot {q}}^{i}=\left\{ {{q}^{i}},H \right\}_{PB},~~{\dot {p}}^{i}=\left\{ {{p}^{i}},H \right\}_{PB},~~i=1,\cdots ,n\]where ${\dot {q}}^{i}=\frac{d}{dt}{{q}^{i}}$.

\subsection{Poisson bracket and Hamilton's equations in complex coordinates}
For certain calculations one sees that it is easier to use Hamilton's equations defined on ${{\mathbb{C}}^{n}}$ rather than ${{\mathbb{R}}^{2n}}$. it means that the PB defined on ${{\mathbb{R}}^{2n}}$ is replaced by  ${{\mathbb{C}}^{n}}$ which is more simple in formulations. One considers the Hamiltonian as a real valued function of the complex variables $z$ and $\overline{z}$ where \cite{3,6}
\begin{equation}\label{eq1}
  {{z}^{j}}={{q}^{j}}+\sqrt{-1}{{p}^{j}},~\overline{{{z}^{j}}}={{q}^{j}}-\sqrt{-1}{{p}^{j}},~~j=1,\cdots ,n
\end{equation}
where partial derivatives are related through the following expressions
\begin{equation}\label{eq2}
\frac{\partial }{\partial {{z}^{j}}}=\frac{1}{2}\left( \frac{\partial }{\partial {{q}^{j}}}-\sqrt{-1}\frac{\partial }{\partial {{p}^{j}}} \right),~~\frac{\partial }{\partial \overline{{{z}^{j}}}}=\frac{1}{2}\left( \frac{\partial }{\partial {{q}^{j}}}+\sqrt{-1}\frac{\partial }{\partial {{p}^{j}}} \right)
\end{equation}
The Poisson bracket (PB) defined on ${{\mathbb{C}}^{n}}$ in complex coordinates of two real valued ${{C}^{1}}$ functions of $z$ and $\overline{z}$ takes the following form \cite{3,6}
\begin{equation}\label{eq4}
  \left\{f,H \right\}_{PB}=2\sqrt{-1}\sum\limits_{j}{\left( \frac{\partial f}{\partial \overline{{{z}^{j}}}}\frac{\partial H}{\partial {{z}^{j}}}-\frac{\partial f}{\partial {{z}^{j}}}\frac{\partial H}{\partial \overline{{{z}^{j}}}} \right)}
\end{equation}
This can be verified from Eq \eqref{eq3} by using Eq \eqref{eq1} and Eq \eqref{eq2}. Hamilton's equations (simplified as HE) in complex coordinates then take the form \cite{3}
\begin{equation}\label{eq7}
  {\dot {z}}^{j}=\frac{d{{z}^{j}}}{dt}=\left\{ {{z}^{j}},H \right\}_{PB}=-2\sqrt{-1}\frac{\partial H}{\partial \overline{{{z}^{j}}}},~~j=1,\cdots ,n
\end{equation}In fact, we also get evolution equation in terms of $\overline{{{z}^{j}}}$,
$$\frac{d}{dt}\overline{{{z}^{j}}}={{\left\{ \overline{{{z}^{j}}},H \right\}}_{PB}}=2\sqrt{-1}\frac{\partial H}{\partial {{z}^{j}}}$$
Plugging them to \eqref{eq4}, and we obtain
\begin{align}
  \frac{df}{dt}&= {{\left\{ f,H \right\}}_{PB}}=2\sqrt{-1}\sum\limits_{j}{\left( \frac{\partial f}{\partial \overline{{{z}^{j}}}}\frac{\partial H}{\partial {{z}^{j}}}-\frac{\partial f}{\partial {{z}^{j}}}\frac{\partial H}{\partial \overline{{{z}^{j}}}} \right)}  \notag\\
 & =2\sqrt{-1}\sum\limits_{j}{\frac{\partial f}{\partial \overline{{{z}^{j}}}}\frac{\partial H}{\partial {{z}^{j}}}-2\sqrt{-1}\sum\limits_{j}{\frac{\partial H}{\partial \overline{{{z}^{j}}}}\frac{\partial f}{\partial {{z}^{j}}}}}  \notag\\
 & =\sum\limits_{j}{\frac{d\overline{{{z}^{j}}}}{dt}\frac{\partial f}{\partial \overline{{{z}^{j}}}}+\sum\limits_{j}{\frac{d{{z}^{j}}}{dt}\frac{\partial f}{\partial {{z}^{j}}}}} \notag
\end{align}
It derives ordinary time operator $$\frac{d}{dt}={{\left\{\cdot ,H \right\}}_{PB}}=\sum\limits_{j}{\left( \frac{d\overline{{{z}^{j}}}}{dt}\frac{\partial }{\partial \overline{{{z}^{j}}}}+\frac{d{{z}^{j}}}{dt}\frac{\partial }{\partial {{z}^{j}}} \right)}$$
where $\frac{d{{z}^{j}}}{dt}, \frac{d}{dt}\overline{{{z}^{j}}}$ are given by
using Poisson bracket (PB) in the form as \eqref{eq7} shown.

Note that PB \eqref{eq4} defined on ${{\mathbb{C}}^{n}}$ satisfies the non-degeneracy, that is,
if $f$ in the ${{\left\{ f,g \right\}}_{PB}}=0$  holds for all smooth function $g$, then $f=Const$.

\section{The GSPB and geometric bracket in real coordinates}
In this section, we will briefly review the entire theoretical framework of generalized covariant Hamiltonian system defined by the generalized structural Poisson bracket totally based on the paper \cite{5} as a revision of generalized Poisson bracket.

To begin with the generalized Poisson bracket (GPB) \cite{4,5} which
is defined on ${{\mathbb{R}}^{r}}$ as the bilinear operation
\[{{\left\{ f,g \right\}}_{GPB}}={{\nabla }^{T}}fJ\nabla g={{J}_{ij}}\frac{\partial f}{\partial {{x}_{i}}}\frac{\partial g}{\partial {{x}_{j}}}\] where structural matrix $J$ satisfies antisymmetric ${{J}_{ij}}={{\left\{ {{x}_{i}},{{x}_{j}} \right\}}_{GPB}}=-{{J}_{ji}}$. Generalized Hamiltonian system (GHS) is defined as \cite{5}
\[{\dot{x}}=\frac{dx}{dt}=J\left( x \right)\nabla H\left( x \right),~~~x\in {{\mathbb{R}}^{m}}\] The GHS also can be given by using generalized Poisson bracket (GPB)
\[{\dot{x_{i}}}=\frac{dx_{i}}{dt}=\left\{ {{x}_{i}},H \right\}_{GPB}={{{J}_{ij}}\frac{\partial H}{\partial {{x}_{j}}}},~~~x\in {{\mathbb{R}}^{m}}\]
Notice that the only difference between the PB defined on ${{\mathbb{R}}^{2n}}$ and GPB on ${{\mathbb{R}}^{r}}$ is whether there exists non-degeneracy or not in definition, the former one has it while the latter doesn't have such restriction. In other words, the GPB is built on the zero restriction of the non-degeneracy.

Let $M$ be a smooth manifold and let $s=s(x)$ be a smooth real structural function on $M$ which is completely determined by the structure of manifold $M$.
Without loss of generality, we give a GSPB in a abstract covariant form,
\begin{definition}\label{d2} \cite{5}
  The GSPB of two functions $f,g\in {{C}^{\infty }}\left( M,\mathbb{R} \right)$ on $M$ is defined as
  $$\left\{ f,g \right\}={{\left\{ f,g \right\}}_{GPB}}+G\left(s, f,g \right)$$
  where $G\left( s,f,g \right)=-G\left(s, g,f \right)$ is called geometric bracket.
\end{definition}
It is remarkable to see that the GSPB representation naturally admits a dynamical geometric bracket formula, basically.  In this abstract representation of the generalized structural Poisson bracket, we see that geobracket equals
\[
  G\left(s, f,g \right)= \left\{ f,g \right\}-{{\left\{ f,g \right\}}_{GPB}},~~f,g\in {{C}^{\infty }}\left( M,\mathbb{R} \right)
\]
which always satisfies the covariant condition $G\left(s, f,g \right)\neq 0$.

\begin{theorem}[geometric bracket]\cite{5}
The geobracket is expressed as
 \begin{align}
 G\left(s,f,g \right) =f{{\left\{ s ,g \right\}}_{GPB}}-g{{\left\{ s ,f \right\}}_{GPB}} \notag
\end{align}for two functions $f,g\in {{C}^{\infty }}\left( M,\mathbb{R} \right)$ on $M$.
\end{theorem}
Intuitively, one can understand this defined formulation by noting that the second part, geobracket, can be treated as the second part of the geodesic equation, and in so doing will make it clear. The GSPB depends smoothly on both ${{\left\{ f,g \right\}}_{GPB}}$ and the geobracket $G\left(s,f,g \right) $. Obviously, the geobracket $G\left(s,f,g \right) $ is necessary for a complete Hamiltonian syatem, as a result of the geobracket $G\left(s,f,g \right) $, it can be generally used to depict nonlinear system in a non-Euclidean space, and for general manifolds.   As GSPB defined, \begin{align}
\left\{ f,g \right\}  &={{\left\{ f,g \right\}}_{GPB}}+G\left( s,f,g \right) \notag\\
 & ={{\left\{ f,g \right\}}_{GPB}}+f{{\left\{ s,g \right\}}_{GPB}}-g{{\left\{ s,f \right\}}_{GPB}} \notag
\end{align}
The GCHS is completely assured and determined by the GSPB. Therefore, the GCHS can be distinctly obtained as follows.

\subsection{The GCHS in real coordinates}
\begin{theorem}[GCHS]\cite{5}
The GCHS of two functions $f,H\in {{C}^{\infty }}\left( M,\mathbb{R} \right)$ on $M$ is defined as
\[ \frac{\mathcal{D}f}{dt}=\left\{ f,H \right\}={{\left\{ f,H \right\}}_{GPB}}+G\left(s,f,H \right) \]for ${{\mathbb{R}}^{r}}$,  the geobracket is
\begin{align}
 G\left(s,f,H \right) =f{{\left\{ s ,H \right\}}_{GPB}}-H{{\left\{ s ,f \right\}}_{GPB}} \notag
\end{align}in terms of $f$ and Hamiltonian $H$,  $s$ is structural function on $M$.
\end{theorem}In fact, general covariance of the GCHS holds naturally.
The GCHS is totally dependent on the GSPB and accordingly defined.

\begin{theorem}[TGHS,S-dynamics, GCHS]\label{t5}\cite{5,7}
The TGHS, S-dynamics, GCHS can be respectively formulated as
\begin{description}
\item[TGHS:] $\frac{df}{dt}= {\dot {f}}={{\left\{ f,H \right\}}_{GPB}}-H{{\left\{ s ,f \right\}}_{GPB}}$.
  \item[S-dynamics:] $\frac{ds }{dt}=w={{\left\{ s ,H \right\}}_{GPB}}=\left\{ 1,H \right\}$.
  \item[GCHS:] $\frac{\mathcal{D}f}{dt}=\left\{ f,H \right\}={{\left\{ f,H \right\}}_{GPB}}+G\left(s,f,H \right)$.
\end{description}
\end{theorem}
By defining the GSPB which can basically solve and describe an entire Hamiltonian system, even nonlinear system. Thus, in order to state this point. To start with, we recall the basic notions of the GSPB. We explain that the GCHS corresponds to non-Euclidean space which corresponds to curved coordinate systems in curved space-time. The geobracket $G\left(s,f,H \right)$ is the part as a correction term for the general covariance which certainly means nonlinear system.

In the following, the covariant equilibrium equation of the GCHS as a special case is naturally generated as follows.
\begin{corollary}\cite{5,7}
 The covariant equilibrium equation of the GCHS is $\frac{\mathcal{D}f}{dt}=\left\{ f,H \right\}=0$, i.e, $${{\left\{ f,H \right\}}_{GPB}}+G\left(s,f,H \right)=0$$holds, then $f$ is called covariant
conserved quantity.
\end{corollary}

According to the GCHS in terms of the momentum, we can deduce the generalized force field.
\begin{corollary}\label{c3}\cite{7}
The generalized force field $F$ on the generalized Poisson manifold $\left( P,S,\left\{ \cdot ,\cdot  \right\} \right)$ is shown as  $$F=\frac{d}{dt}p=-DH$$ Its components is $F_{k}={\dot{p}_{k}}=-{{D}_{k}}H$.
\end{corollary}

\begin{theorem}\cite{5,7}
There exists two different canonical Hamilton equations on manifolds respectively given by
  \begin{description}
    \item[Canonical thorough Hamilton equations] \[\frac{d{{x}_{k}}}{dt}={\dot{x}_{k}}={{J}_{jk}}F_{j},~~\frac{d{{p}_{k}}}{dt}={\dot{p}_{k}}=F_{k}\]

    \item[Canonical covariant Hamilton equations]
  \[\frac{\mathcal{D}{{x}_{k}}}{dt}=\left\{{{x}_{k}} ,H \right\},~~\frac{\mathcal{D}{{p}_{k}}}{dt}=\left\{{{p}_{k}} ,H\right\}\]
  \end{description}
where $\left\{ \cdot ,~\cdot  \right\}={{\left\{ \cdot ,~\cdot  \right\}}_{GPB}}+G\left(s,\cdot ,~\cdot  \right)$ is the GSPB.
\end{theorem}In \cite{5}, we have introduced an geometric approach to solve Hamiltonian mechanical problems based on the generalized structural Poisson bracket formulation.
Essentially, there always exists a structural function $s$ such that geometric bracket $G\left(s, f,g \right)$ holds for the GSPB and the GCHS on manifolds.

\begin{corollary}\label{c2}\cite{7}
The ordinary time derivative can be rewritten in a form
$d/dt={\dot{x}_{i}}{{\partial }_{i}}$ and covariant time derivative is of the form $\mathcal{D}/dt={\dot{x}_{i}}{{D }_{i}}$,
where ${\dot{x}_{i}}$ is the TGHS in terms of ${x}_{i}$,
and then for the generalized force field $F$, we get generalized force field expressed by
${{F}_{k}}={\dot{x}_{i}}{{\xi }_{ik}}$ in terms of the momentum.
Similarly, for the S-dynamics, it yields
$w={\dot{x}_{i}}{{\partial }_{i}}s={\dot{x}_{i}}{{A}_{i}}$
in terms of the structure function $s$,
and the TGHS is given by
$df/dt={\dot{x}_{i}}{{\partial }_{i}}f$ for any function $f$.
\end{corollary}
Obviously, we see that the theorem \ref{t5} can be rewritten as the following form by using corollary \ref{c2}.
\begin{theorem}[TGHS,S-dynamics, GCHS]\label{t6}
The TGHS, S-dynamics, GCHS can be respectively formulated as
\begin{description}
\item[TGHS:] $\frac{df}{dt}= {\dot {f}}={\dot{x}_{i}}{{\partial }_{i}}f$.
  \item[S-dynamics:] $\frac{ds }{dt}=w={\dot{x}_{i}}{{\partial }_{i}}s$.
  \item[GCHS:] $\frac{\mathcal{D}f}{dt}=\left\{ f,H \right\}={\dot{x}_{i}}{{D }_{i}}f$.
\end{description}where ${\dot{x}_{i}}={{J}_{ji}}F_{j}$ is the TGHS in terms of ${x}_{i}$.
\end{theorem}

There is a very important theorem about the geometrio induced by the  structural operator $\widehat{S}$ and structure derivative is given by
\begin{theorem}[Geometrio]\label{t4}\cite{7}
The structural operator $\widehat{S}$ can induce the following geometrio
 \[\hat{S}{{\left( {{x}_{k}},{{p}_{i}},H \right)}^{T}}={{\left( {{b}_{k}},{{A}_{i}},w \right)}^{T}}\]
in terms of position ${x}_{k}$, momentum ${p}_{i}$ and Hamiltonian $H$ respectively.
\end{theorem}
Note that the entire framework of the GCHS and GSPB can be rewritten in another form by rising lower index of each parameter such as ${{x}_{i}}\to {{x}^{i}}$, the conclusions remain the same.

\section{The GSPB in complex coordinates}
Let $M$ be a smooth manifold and let $s=s\left( z,\overline{z} \right)$ be a smooth complex structural function in terms of complex coordinates $z,\overline{z}$ on $M$ which is completely determined by the structure of manifold $M$, where $z=\left( {{z}^{1}},\cdots ,{{z}^{j}} \right)$ has been used.  Note that $s$ in complex situation may take a different form from the real situation, the exact form is taken according to the certain question we study.
Without loss of generality,  we give the definition of generalized structural Poisson bracket of a abstract covariant form in complex coordinates.

\subsection{The GSPB and geometric bracket in complex coordinates}
In this section, firstly, let's abandon the non-degenerate condition of Poisson bracket \eqref{eq4} just like the GPB defined on such condition, secondly, it generalizes the Poisson bracket $$\left\{ f,g \right\}_{PB}=2\sqrt{-1}\sum\limits_{j}{\left( \frac{\partial f}{\partial \overline{{{z}^{j}}}}\frac{\partial g}{\partial {{z}^{j}}}-\frac{\partial f}{\partial {{z}^{j}}}\frac{\partial g}{\partial \overline{{{z}^{j}}}} \right)}$$ defined on ${{\mathbb{C}}^{n}}$ to the GSPB defined on ${{\mathbb{C}}^{r}}$ without the restriction of non-degeneracy in complex coordinates by considering the structural function only relied on the manifold itself. Here we always ask GSPB in complex coordinates to hold without such restriction of non-degeneracy.

\begin{definition}\label{d1}
The GSPB of two functions $f,g\in {{C}^{\infty }}\left( M \right)$ in complex coordinates is defined as
\[\left\{ f,g \right\}={{\left\{ f,g \right\}}_{PB}}+G\left(s, f,g \right)\]
and the geometric bracket is $$G\left(s,f,g \right) =f{{\left\{ s ,g \right\}}_{PB}}-g{{\left\{ s ,f \right\}}_{PB}}$$
for two functions $f,g\in {{C}^{\infty }}\left( M,\mathbb{C} \right)$ on $M$.

\end{definition}
Note that smooth complex structural function $s$ here is a function in phase space in terms of complex variable ${z}^{j}$, namely, $s=s\left( {{z}^{j}} \right)$.
In this abstract representation of the generalized structural Poisson bracket, we see that geobracket is equal to
\[G\left(s, f,g \right)= \left\{ f,g \right\}-{{\left\{ f,g \right\}}_{PB}},~~f,g\in {{C}^{\infty }}\left( M\right)\]
which always satisfies the covariant condition $s\left( f,g \right)\neq 0$.
As defined above, we know that there always exists the geobracket $G\left( s,f,g \right)\neq 0$ such that the generalized structural Poisson bracket holds on manifolds.

The non-zero of the geobracket clearly embodies the unique structure of the manifolds $M$, it's for an invariant covariance, definitely.
 Note that the geobracket satisfies the antisymmetry, that is, $$G\left(s, f,g \right)=-G\left(s, g,f \right)$$ and one can easily verify that  $\left\{ \cdot,\cdot \right\}$ is skew-symmetric and bilinear. This preserves the classical properties. Meanwhile, it has some other properties such as linearity
\[G\left(s, \lambda f+\mu g,h \right)=\lambda G\left(s, f,h \right)+\mu G\left(s, g,h \right)\]
The GSPB of two functions $f,g\in {{C}^{\infty }}\left( M \right)$ in complex coordinates also can be expressed as 
\begin{align}
  \left\{ f,g \right\}& ={{\left\{ f,g \right\}}_{PB}}+f{{\left\{ s,g \right\}}_{PB}}-g{{\left\{ s,f \right\}}_{PB}}  \notag\\
 & =\Omega \left( {{X}_{f}},{{X}_{g}} \right)+f\Omega \left( {{X}_{s}},{{X}_{g}} \right)-g\Omega \left( {{X}_{s}},{{X}_{f}} \right) \notag \\
 & =\left\langle {{X}_{f}},J{{X}_{g}} \right\rangle +f\left\langle {{X}_{s}},J{{X}_{g}} \right\rangle -g\left\langle {{X}_{s}},J{{X}_{f}} \right\rangle   \notag
\end{align}
where \[{{\left\{ f,g \right\}}_{PB}}=\Omega \left( {{X}_{f}},{{X}_{g}} \right)=\left\langle {{X}_{f}},J{{X}_{g}} \right\rangle \]
Essentially, the completeness of the GSPB can be seen from its formula above, namely, it's amazingly given by the geometric bracket.

\subsection{Generalized partial derivatives}
In this section, one will generalize the partial derivatives Eq \eqref{eq2} to more general form based on Eq \eqref{eq1}.  The generalized method one takes is to introduce the structural function $s$. Hence, the transformation of partial derivatives Eq \eqref{eq2} is given by the following general form
\begin{align}\label{eq6}
  & \frac{\partial }{\partial {{z}^{j}}}\to \frac{\rm{D}}{\partial {{z}^{j}}}=\frac{\partial }{\partial {{z}^{j}}}+\frac{\partial s}{\partial {{z}^{j}}} \\
 & \frac{\partial }{\partial \overline{{{z}^{j}}}}\to \frac{\rm{D}}{\partial \overline{{{z}^{j}}}}=\frac{\partial }{\partial \overline{{{z}^{j}}}}+\frac{\partial s}{\partial \overline{{{z}^{j}}}} \notag
\end{align}
where structural partial derivatives in terms of the complex variables $z$ and $\overline{z}$ are shown as \[\frac{\partial s}{\partial {{z}^{j}}}=\frac{1}{2}\left( \frac{\partial s}{\partial {{q}^{j}}}-\sqrt{-1}\frac{\partial s}{\partial {{p}^{j}}} \right),~~\frac{\partial s}{\partial \overline{{{z}^{j}}}}=\frac{1}{2}\left( \frac{\partial s}{\partial {{q}^{j}}}+\sqrt{-1}\frac{\partial s}{\partial {{p}^{j}}} \right)\]in which structural function $s$ can be taken as a real function only associated with domain, $\frac{\partial s}{\partial {{z}^{j}}},~\frac{\partial s }{\partial \overline{{{z}^{j}}}}$ can be called as structural partial derivative and $\frac{\partial s}{\partial {{z}^{j}}},~\frac{\partial s}{\partial \overline{{{z}^{j}}}}\ne 0$.

As Eq \eqref{eq4} defined,  one also takes the same procedure to obtain the GSPB, simplified as abbreviation GSPB in complex coordinates. In order to obtain the GSPB, the method is using generalized partial derivative Eq \eqref{eq6} $\frac{\rm{D}}{\partial {{z}^{j}}},~\frac{\rm{D}}{\partial \overline{{{z}^{j}}}}$ to replace the ordinary partial derivative Eq \eqref{eq2} $\frac{\partial }{\partial {{z}^{j}}},\frac{\partial }{\partial \overline{{{z}^{j}}}}$.  As a consequence, the GSPB of two real valued ${{C}^{1}}$ functions of $z$ and $\overline{z}$ takes the following form in complex coordinates
\begin{equation}\label{eq5}
  {{\left\{f,H \right\}}_{PB}}\to \left\{f,H \right\}=2\sqrt{-1}\sum\limits_{j}{\left( \frac{{\rm{D}}f}{\partial \overline{{{z}^{j}}}}\frac{{\rm{D}}H}{\partial {{z}^{j}}}-\frac{{\rm{D}}f}{\partial {{z}^{j}}}\frac{{\rm{D}}H}{\partial \overline{{{z}^{j}}}} \right)}
\end{equation}
For specific calculations, it follows
\begin{align}
  & \frac{{\rm{D}}f}{\partial \overline{{{z}^{j}}}}\frac{{\rm{D}}H}{\partial {{z}^{j}}}-\frac{{\rm{D}}f}{\partial {{z}^{j}}}\frac{{\rm{D}}H}{\partial \overline{{{z}^{j}}}} \notag\\
 & =\frac{\partial f}{\partial \overline{{{z}^{j}}}}\frac{\partial H}{\partial {{z}^{j}}}-\frac{\partial F}{\partial {{z}^{j}}}\frac{\partial H}{\partial \overline{{{z}^{j}}}}+f\left( \frac{\partial s}{\partial \overline{{{z}^{j}}}}\frac{\partial H}{\partial {{z}^{j}}}-\frac{\partial s}{\partial {{z}^{j}}}\frac{\partial H}{\partial \overline{{{z}^{j}}}} \right) \notag\\
 & \begin{matrix}
   {} & {} & {}  \\
\end{matrix}+H\left( \frac{\partial f}{\partial \overline{{{z}^{j}}}}\frac{\partial s}{\partial {{z}^{j}}}-\frac{\partial f}{\partial {{z}^{j}}}\frac{\partial s}{\partial \overline{{{z}^{j}}}} \right)+fH\left( \frac{\partial s}{\partial \overline{{{z}^{j}}}}\frac{\partial s}{\partial {{z}^{j}}}-\frac{\partial s}{\partial {{z}^{j}}}\frac{\partial s}{\partial \overline{{{z}^{j}}}} \right) \notag
\end{align}
And then Eq \eqref{eq5} is certainly shown as
\begin{align}
\left\{f,H \right\}  & =2\sqrt{-1}\sum\limits_{j}{\left( \frac{\partial f}{\partial \overline{{{z}^{j}}}}\frac{\partial H}{\partial {{z}^{j}}}-\frac{\partial f}{\partial {{z}^{j}}}\frac{\partial H}{\partial \overline{{{z}^{j}}}} \right)}+2\sqrt{-1}f\sum\limits_{j}{\left( \frac{\partial s}{\partial \overline{{{z}^{j}}}}\frac{\partial H}{\partial {{z}^{j}}}-\frac{\partial s}{\partial {{z}^{j}}}\frac{\partial H}{\partial \overline{{{z}^{j}}}} \right)} \notag\\
 & +2\sqrt{-1}H\sum\limits_{j}{\left( \frac{\partial f}{\partial \overline{{{z}^{j}}}}\frac{\partial s}{\partial {{z}^{j}}}-\frac{\partial f}{\partial {{z}^{j}}}\frac{\partial s}{\partial \overline{{{z}^{j}}}} \right)}+2\sqrt{-1}fH\sum\limits_{j}{\left( \frac{\partial s}{\partial \overline{{{z}^{j}}}}\frac{\partial s}{\partial {{z}^{j}}}-\frac{\partial s}{\partial {{z}^{j}}}\frac{\partial s}{\partial \overline{{{z}^{j}}}} \right)} \notag\\
 & ={{\left\{f,H \right\}}_{PB}}+f{{\left\{ s,H \right\}}_{PB}}-H{{\left\{ s,f \right\}}_{PB}} \notag
\end{align}
where one has used the identity ${{\left\{ s,s \right\}}_{PB}}=2\sqrt{-1}\sum\limits_{j}{\left( \frac{\partial s}{\partial \overline{{{z}^{j}}}}\frac{\partial s}{\partial {{z}^{j}}}-\frac{\partial s}{\partial {{z}^{j}}}\frac{\partial s}{\partial \overline{{{z}^{j}}}} \right)}=0$, and antisymmetric property ${{\left\{ s,f \right\}}_{PB}}=-{{\left\{f,s \right\}}_{PB}}$ to deduce the GSPB based on the transformation Eq \eqref{eq6}.

\begin{theorem}[GCHS]\label{t2}
  The GCHS in complex coordinates based on the definition \ref{d1} is
  \begin{equation}\label{eq8}
    \left\{f,H \right\}={{\left\{ f,H \right\}}_{PB}}+G\left( s, f,H \right)
  \end{equation}
where the geobracket is
\begin{equation}\label{eq13}
  G\left( s,f,H \right)=f{{\left\{ s,H \right\}}_{PB}}-H{{\left\{ s,f \right\}}_{PB}}
\end{equation}
\end{theorem} One has followed the steps described above for constructing a new and invariant bracket for the more general Hamiltonian system.
It is clear that the GSPB satisfies the antisymmetric $\left\{f,H \right\}=-\left\{ H,f \right\}$. Notice that the GSPB is a combination of three parts, the PB ${{\left\{f,H \right\}}_{PB}}$ as known classic Hamilton system defined, the other two parts $f{{\left\{ s,H \right\}}_{PB}}$ and $H{{\left\{ s,f\right\}}_{PB}}$ naturally emerged together in the geobracket Eq \eqref{eq13} are induced by the structural function $s$.  By analyzing the Eq \eqref{eq8}, it appears that the GSPB can be divided into two parts:
\[\left\{f,H \right\}=\underbrace{{{\left\{f,H \right\}}_{PB}}-H{{\left\{ s,f \right\}}_{PB}}}_{{\dot {f}}}+f\underbrace{{{\left\{ s,H \right\}}_{PB}}}_{w}\]In other words,  the framework of the GSPB contains two different dynamics which are connected to the function $s$ and its derivative operation
\[\frac{df}{dt}={\dot {f}}={{\left\{f,H \right\}}_{PB}}-H{{\left\{ s,f \right\}}_{PB}},~~w={{\left\{ s,H \right\}}_{PB}}\]
In particular, the latter dynamics has no connections with function $f$, it only relies on the structure of domain one studies, namely, it's independent of $f$. Based on the \eqref{eq18}, the GCHS in complex coordinates can be expressed as
\begin{align}
 \left\{ f,H \right\} &={{\left\{ f,H \right\}}_{PB}}+f{{\left\{ s,H \right\}}_{PB}}-H{{\left\{ s,f \right\}}_{PB}} \notag\\
 & =\Omega \left( {{X}_{f}},{{X}_{H}} \right)+f\Omega \left( {{X}_{s}},{{X}_{H}} \right)-H\Omega \left( {{X}_{s}},{{X}_{f}} \right) \notag\\
 & =\left\langle {{X}_{f}},J{{X}_{H}} \right\rangle +f\left\langle {{X}_{s}},J{{X}_{H}} \right\rangle -H\left\langle {{X}_{s}},J{{X}_{f}} \right\rangle  \notag
\end{align}
where S-dynamics and TGHS are separately written as
\begin{align}
  & w=\left\langle {{X}_{s}},J{{X}_{H}} \right\rangle  \notag \\
 & df/dt=\left\langle {{X}_{f}},J{{X}_{H}} \right\rangle -H\left\langle {{X}_{s}},J{{X}_{f}} \right\rangle   \notag
\end{align}
In the following, the covariant equilibrium equation of the GCHS as a special case is naturally generated as follows.
\begin{corollary}\label{c4}
 The covariant equilibrium equation of the GCHS in complex coordinates is $\frac{\mathcal{D}f}{dt}=\left\{ f,H \right\}=0$ such that $${{\left\{ f,H \right\}}_{PB}}+G\left( s ;f,H \right)=0$$holds, then $f$ is called covariant conserved quantity.
\end{corollary}

\subsection{GSPB in complex form between complex variables}
Let's do a calculate between ${{z}^{j}},\overline{{{z}^{j}}}$ by using the PB,
\[{{\left\{ {{z}^{j}},\overline{{{z}^{j}}} \right\}}_{PB}}=2\sqrt{-1}\sum\limits_{j}{\left( \frac{\partial {{z}^{j}}}{\partial \overline{{{z}^{j}}}}\frac{\partial \overline{{{z}^{j}}}}{\partial {{z}^{j}}}-\frac{\partial \overline{{{z}^{j}}}}{\partial \overline{{{z}^{j}}}}\frac{\partial {{z}^{j}}}{\partial {{z}^{j}}} \right)}=-2\sqrt{-1}\]
As a comparison, the same computation goes to the ${{z}^{j}},\overline{{{z}^{j}}}$ by using the GSPB, let's see how it becomes
\[\left\{ {{z}^{j}},\overline{{{z}^{j}}} \right\}={{\left\{ {{z}^{j}},\overline{{{z}^{j}}} \right\}}_{PB}}+G\left( s,{{z}^{j}},\overline{{{z}^{j}}} \right)\]
Hence, we just need to evaluate the geometric bracket \[G\left( s,{{z}^{j}},\overline{{{z}^{j}}} \right)={{z}^{j}}{{\left\{ s,\overline{{{z}^{j}}} \right\}}_{PB}}-\overline{{{z}^{j}}}{{\left\{ s,{{z}^{j}} \right\}}_{PB}}\] with respect to ${{z}^{j}},\overline{{{z}^{j}}}$, then
\[{{\left\{ s,{{z}^{j}} \right\}}_{PB}}=2\sqrt{-1}\frac{\partial s}{\partial \overline{{{z}^{j}}}},~~{{\left\{ s,\overline{{{z}^{j}}} \right\}}_{PB}}=-2\sqrt{-1}\frac{\partial s}{\partial {{z}^{j}}}\]
Therefore, the geometric bracket in details is given by
\begin{align}
 G\left( s,{{z}^{j}},\overline{{{z}^{j}}} \right) & ={{z}^{j}}{{\left\{ s,\overline{{{z}^{j}}} \right\}}_{PB}}-\overline{{{z}^{j}}}{{\left\{ s,{{z}^{j}} \right\}}_{PB}} \notag\\
 & =-2\sqrt{-1}\left( {{z}^{j}}\frac{\partial s}{\partial {{z}^{j}}}+\overline{{{z}^{j}}}\frac{\partial s}{\partial \overline{{{z}^{j}}}} \right) \notag
\end{align}
Conclusively, the GSPB in complex form in terms of  ${{z}^{j}},\overline{{{z}^{j}}}$ is expressed as
\[\left\{ {{z}^{j}},\overline{{{z}^{j}}} \right\}=-2\sqrt{-1}\left( 1+{{z}^{j}}\frac{\partial s}{\partial {{z}^{j}}}+\overline{{{z}^{j}}}\frac{\partial s}{\partial \overline{{{z}^{j}}}} \right)\]
and \begin{align}
  & \left\{ {{z}^{j}},{{z}^{j}} \right\}=0 \notag\\
 & \left\{ \overline{{{z}^{j}}},\overline{{{z}^{j}}} \right\}=0 \notag
\end{align}

\section{The GCHS in complex coordinates}
This section analyzes and defines the S-dynamics and TGHS in complex coordinates, expounds the covariant mechanism of GCHS, plus, it excogitates a methods which could accurately evaluate the stability of the GCHS.

In the framework of GSPB, then the generalized covariant Hamilton's equations in complex coordinates take the form\[\left\{ {{z}^{j}},H \right\}={{\left\{ {{z}^{j}},H \right\}}_{PB}}-H{{\left\{ s,{{z}^{j}} \right\}}_{PB}}+{{z}^{j}}{{\left\{ s,H \right\}}_{PB}}\]
then accordingly Eq \eqref{eq7} is completely rewritten as
\begin{equation}\label{eq11}
  {\dot {z}}^{j}={{\left\{ {{z}^{j}},H \right\}}_{PB}}-H{{\left\{ s,{{z}^{j}} \right\}}_{PB}}=-2\sqrt{-1}\frac{{\rm{D}}H}{\partial \overline{{{z}^{j}}}},~~j=1,\cdots ,n
\end{equation}
This is a correction for Eq \eqref{eq7}, obviously, one regards it as the thorough generalized Hamiltonian system (TGHS), then GCHS in complex coordinates is basically given by
\[\left\{ {{z}^{j}},H \right\}={\dot {z}}^{j}+{{z}^{j}}{{\left\{ s,H \right\}}_{PB}},~~j=1,\cdots ,n\]
If one defines S-dynamics as $w={{\left\{ s,H \right\}}_{PB}}$, then $\left\{ {{z}^{j}},H \right\}={\dot {z}}^{j}+{{z}^{j}}w$, or in the form
\begin{equation}\label{eq9}
  \left\{ {{z}^{j}},H \right\}=-2\sqrt{-1}\frac{{\rm{D}}H}{\partial \overline{{{z}^{j}}}}+{{z}^{j}}{{\left\{ s,H \right\}}_{PB}}=-2\sqrt{-1}\frac{{\rm{D}}H}{\partial \overline{{{z}^{j}}}}+{{z}^{j}}w
\end{equation}
Therefore, this is the most complete equation form for Hamiltonian system in reality. One bracket which gets two dynamic subsystem involved in a whole equation for describing one system, it has covariant property under coordinates transformation.

As a consequence of theorem \ref{t2}, let's write it in terms of ${z}^{j}$.
\begin{theorem}[TGHS,S-dynamics, GCHS]\label{t1}
The TGHS, S-dynamics, GCHS in complex coordinates can be respectively formulated as
\begin{description}
\item[TGHS:] $\frac{d{{z}^{j}}}{dt}={\dot {z}}^{j}={{\left\{ {{z}^{j}},H \right\}}_{PB}}-H{{\left\{ s,{{z}^{j}} \right\}}_{PB}}=-2\sqrt{-1}\frac{{\rm{D}}H}{\partial \overline{{{z}^{j}}}}$.
  \item[S-dynamics:] $w={{\left\{ s,H \right\}}_{PB}}=\left\{1,H \right\}$.
  \item[GCHS:] $\frac{\mathcal{D}{{z}^{j}}}{dt}=\left\{ {{z}^{j}},H \right\}={\dot {z}}^{j}+{{z}^{j}}w$.
\end{description}where $\frac{\mathcal{D}}{dt}=\frac{d}{dt}+w$ is the covariant time operator, and $\left\{\cdot,\cdot\right\}$ is the GSPB in complex coordinates.
\end{theorem}
As revised theorem \ref{t1} shown, one has improved and perfected Hamilton's equations and the Poisson bracket. Fundamentally, these covariant expressions as complete form remedy the deficiency of Hamilton's equations and the Poisson bracket, it's suitable for the general case including the manifold equipped with different structure, in other words, non-Euclidean space. Actually, these covariant form of Hamilton system broke through many of the boundedness.  Thus, one can easily conclude that a rounded Hamiltonian system should be attached to the structural function $s$ of the domain or the manifolds.
The CHS and GHS together which reveal incomplete defect that all lack two important parts, that is to say, $-H{{\left\{ s,{{z}^{j}} \right\}}_{PB}}$ and S-dynamics respectively.

For the TGHS given by the theorem \ref{t1},  clearly,
$$\frac{d{{z}^{j}}}{dt}=-2\sqrt{-1}\frac{{\rm D}H}{\partial \overline{{{z}^{j}}}}=-2\sqrt{-1}\frac{\partial H}{\partial \overline{{{z}^{j}}}}-2\sqrt{-1}H\frac{\partial s}{\partial \overline{{{z}^{j}}}}$$
By contrast,
$$\frac{d{{z}^{j}}}{dt}={\dot {z}}^{j}={{\left\{ {{z}^{j}},H \right\}}_{PB}}-H{{\left\{ s,{{z}^{j}} \right\}}_{PB}}$$
Obviously, it turns out that $${{\left\{ s,{{z}^{j}} \right\}}_{PB}}=2\sqrt{-1}\frac{\partial s}{\partial \overline{{{z}^{j}}}}$$
Actually, the S-dynamics can be obtained by using the TGHS in the theorem \ref{t1}, that is,
$\frac{d}{dt}s=w={{\left\{ s,H \right\}}_{PB}}$,

By the process of analysing the TGHS, the facts emerge,
\[\frac{{\rm{D}}H}{\partial \overline{{{z}^{j}}}}=0\]is equivalent to the equation
$\frac{d{{z}^{j}}}{dt}=0$, and ${{\left\{ {{z}^{j}},H \right\}}_{PB}}=H{{\left\{ s,{{z}^{j}} \right\}}_{PB}}$.

Therefore, based on the theorem \ref{t1}, let's turn to the GCHS given by theorem \ref{t2}, if let $G=s$ be given on purpose, then it gives rise to
\begin{equation}\label{eq17}
  \left\{ s,H \right\}={{\left\{ s,H \right\}}_{PB}}+s{{\left\{ s,H \right\}}_{PB}}=w+sw
\end{equation}
and the TGHS can be accordingly obtained
\[\frac{ds}{dt}={{\left\{ s,H \right\}}_{PB}}=w\]Hence, to take covariant equilibrium equation \eqref{eq17} into account, namely,  $\frac{\mathcal{D}s}{dt}=\left\{ s,H \right\}=0$, then $\left( 1+s \right)w=0$, it results in two extreme outcomes $s=-1$ or $w=0$,

Obviously, $\left\{ H,H \right\}=0$ is certainly sufficed, it reveals that $H$ is a covariant conserved quantity.  The TGHS then reads
\[\frac{dH}{dt}=-H{{\left\{ s,H \right\}}_{PB}}=-Hw\]in terms of $H$,
its formal solution is \[H={{H}_{0}}{{e}^{-wt}},~~{{H}_{0}}=H\left( 0 \right)\]
This is clearly a standard dynamic solution that doesn't appear in the classical theory, in other words, the classical theory holds as a special case at $w=0$, exponential change to happen in the new and covariant theory is a intact system.

To see how GCHS in complex coordinates is equivalent to the real coordinates, it's easily calculated by the procedure of corresponding definitions and theories which can prove their equivalence and consistence. Our findings are consistent with paper \cite{5} in real coordinates. Without loss of generality, one can abstractly obtain covariant time operator.
\begin{corollary}\label{c1}
Covariant time operator can be shown in the form
\[\frac{\mathcal{D}}{dt}=\left\{ \cdot ,H \right\}=\frac{d}{dt}+w={{\left\{ \cdot ,H \right\}}_{PB}}-H{{\left\{ s,\cdot  \right\}}_{PB}}+{{\left\{ s,H \right\}}_{PB}}\] where
  \begin{align}
  & \frac{d}{dt}={{\left\{ \cdot ,H \right\}}_{PB}}-H{{\left\{ s,\cdot  \right\}}_{PB}}\notag \\
 & w={{\left\{ s,H \right\}}_{PB}}=\left\{ 1,H \right\} \notag
\end{align}are thorough time operator and S-dynamics, respectively.
\end{corollary}
One can easily see that the S-dynamics in complex coordinates is an independent system induced only by the structure function $s$ from the corollary \ref{c1}. The same things goes as the S-dynamics in real coordinates expressed by \cite{5,7}.

Notice that the GCHS is a complete dynamic system described by a covariant equation Eq \eqref{eq9} $\left\{ {{z}^{j}},H \right\}={\dot {z}}^{j}+{{z}^{j}}w$ that is invariant under coordinates transformation as it means. Let's consider the case of corresponding  covariant  equilibrium, the equation given by Eq \eqref{eq9} holds as $\left\{ {{z}^{j}},H \right\}=0$, or the form such as
\begin{equation}\label{eq10}
  {\dot {z}}^{j}=-{{z}^{j}}w,~~z_{0}^{j}={{z}^{j}}\left( 0 \right)
\end{equation}
where TGHS is shown as Eq \eqref{eq11}. Obviously, Eq \eqref{eq10} is a standard dynamic ODE for a real Hamiltonian dynamics,  if we consider a constant case $w={{w}_{0}}$, then its formal solution is easily obtained and given by
\[{{z}^{j}}=z_{0}^{j}{{e}^{-w_{0}t}},~~z_{0}^{j}={{z}^{j}}\left( 0 \right)\]
This dynamic solution fits most of situations and appears mostly in ODE. Using this solution with initial conditions is good for analyzing the stabilization of the system.   Taking a Taylor expansion of Eq \eqref{eq10} at $w={{w}_{0}}$, we can better understand how important the S-dynamics is,
\begin{align}
{{z}^{j}}  &=z_{0}^{j}{{e}^{-{{w}_{0}}t}}=z_{0}^{j}\left( 1-{{w}_{0}}t+\cdots  \right) \notag\\
 & =z_{0}^{j}-z_{0}^{j}\left( 1-{{e}^{-{{w}_{0}}t}} \right) \notag
\end{align}
Clearly, we can see that the constant solution $z_{0}^{j}$ correspond to the classical result of HS, the second term related to the S-dynamics emerges naturally comes from the complete theory-GCHS, it says our rebuilding of GCHS is correct, basically.

Actually, Eq \eqref{eq10} has another expression shown as
\[{{\left\{ {{z}^{j}},H \right\}}_{PB}}-H{{\left\{ s,{{z}^{j}} \right\}}_{PB}}=-{{z}^{j}}w\]rearranging the order and it gives the equation
\begin{equation}\label{eq12}
  {{\left\{ {{z}^{j}},H \right\}}_{PB}}=H{{\left\{ s,{{z}^{j}} \right\}}_{PB}}-{{z}^{j}}w
\end{equation}
One can see that on the left of equation, it represents the classical Hamiltonian equation, on the right side, it's totally dependent on the structure function $s$ and induced by it. As a matter of fact, Eq \eqref{eq12} is only for the case of non-flat space,  in other words, Eq \eqref{eq12} is a equilibrium equation for the non-Euclidean space, especially, for the manifolds equipped with different connections.

Conversely, if $\left\{ {{z}^{j}},H \right\}\neq0$ holds, let $\left\{ {{z}^{j}},H \right\}=h\left( t,{{z}^{j}} \right)$ be given for a non-equilibrium case, then
\[{\dot {z}}^{j}=-{{z}^{j}}w+h\left( t,{{z}^{j}} \right),~~j=1,\cdots ,n\]
where $h=h\left( t,{{z}^{j}} \right)$ is a disturbing function in the system.

As a consequence of theorem \ref{t2}, let's write the GCHS in terms of $\overline{{{z}^{j}}}$.
\begin{theorem}[TGHS,S-dynamics, GCHS]\label{t8}
The TGHS, S-dynamics, GCHS in complex coordinates can be respectively formulated as
\begin{description}
\item[TGHS:] $\frac{d\overline{{{z}^{j}}}}{dt}=2\sqrt{-1}\frac{{\rm{D}}H}{\partial {{z}^{j}}}$.
  \item[S-dynamics:] $w={{\left\{ s,H \right\}}_{PB}}=\left\{1,H \right\}$.
  \item[GCHS:] $\frac{\mathcal{D}}{dt}\overline{{{z}^{j}}}=\left\{ \overline{{{z}^{j}}},H \right\}=2\sqrt{-1}\frac{{\rm{D}}H}{\partial {{z}^{j}}}+\overline{{{z}^{j}}}w $.
\end{description}where $\frac{\mathcal{D}}{dt}=\frac{d}{dt}+w$ is the covariant time operator, and $\left\{\cdot,\cdot\right\}$ is the GSPB in complex coordinates.
\begin{proof}
According to the theorem \ref{t2},  let $f=\overline{{{z}^{j}}}$ be given for derivations. Then

\begin{align}
 \frac{\mathcal{D}}{dt}\overline{{{z}^{j}}} &=\left\{ \overline{{{z}^{j}}},H \right\}={{\left\{ \overline{{{z}^{j}}},H \right\}}_{PB}}+G\left( s,\overline{{{z}^{j}}},H \right) \notag\\
 & =2\sqrt{-1}\frac{\partial H}{\partial {{z}^{j}}}+2\sqrt{-1}H\frac{\partial s}{\partial {{z}^{j}}}+\overline{{{z}^{j}}}w \notag\\
 & =2\sqrt{-1}\left( \frac{\partial H}{\partial {{z}^{j}}}+H\frac{\partial s}{\partial {{z}^{j}}} \right)+\overline{{{z}^{j}}}w \notag\\
 & =2\sqrt{-1}\frac{{\rm{D}}H}{\partial {{z}^{j}}}+\overline{{{z}^{j}}}w \notag
\end{align}
it implies
$$\frac{d\overline{{{z}^{j}}}}{dt}=2\sqrt{-1}\frac{{\rm{D}}H}{\partial {{z}^{j}}}$$
where S-dynamics is $w={{\left\{ s,H \right\}}_{PB}}=\left\{1,H \right\}$ independent to the $f$.

\end{proof}
\end{theorem}
To substitute the theorem \ref{t1} and  theorem \ref{t8} together into Eq \eqref{eq5}, then we can get the GCHS completely shown by

\begin{align}
 \frac{\mathcal{D}}{dt}f &=\left\{ f,H \right\}=2\sqrt{-1}\sum\limits_{j}{\left( \frac{{\rm{D}}H}{\partial {{z}^{j}}}\frac{{\rm{D}}f}{\partial \overline{{{z}^{j}}}}-\frac{{\rm{D}}f}{\partial {{z}^{j}}}\frac{{\rm{D}}H}{\partial \overline{{{z}^{j}}}} \right)} \notag \\
 & =\sum\limits_{j}{\left( \frac{d\overline{{{z}^{j}}}}{dt}\frac{{\rm{D}}}{\partial \overline{{{z}^{j}}}}+\frac{d{{z}^{j}}}{dt}\frac{{\rm{D}}}{\partial {{z}^{j}}} \right)}f  \notag
\end{align}where $\frac{d{{z}^{j}}}{dt},\frac{d\overline{{{z}^{j}}}}{dt}$ are the TGHS in terms of ${z}^{j}$ and $\overline{{{z}^{j}}}$, respectively.
Thusly, the fact
\begin{equation}\label{eq16}
  \frac{\mathcal{D}}{dt}=\left\{\cdot ,H \right\}=\sum\limits_{j}{\left( \frac{d\overline{{{z}^{j}}}}{dt}\frac{{\rm{D}}}{\partial \overline{{{z}^{j}}}}+\frac{d{{z}^{j}}}{dt}\frac{{\rm{D}}}{\partial {{z}^{j}}} \right)}
\end{equation}
follows. Obviously, the fact
\[\frac{\mathcal{D}}{dt}H=\left\{ H,H \right\}=\sum\limits_{j}{\left( \frac{d\overline{{{z}^{j}}}}{dt}\frac{{\rm{D}}H}{\partial \overline{{{z}^{j}}}}+\frac{d{{z}^{j}}}{dt}\frac{{\rm{D}}H}{\partial {{z}^{j}}} \right)}=0\]holds for below identity
\[\sum\limits_{j}{\frac{d{{z}^{j}}}{dt}\frac{{\rm{D}}H}{\partial {{z}^{j}}}}=-\sum\limits_{j}{\frac{d\overline{{{z}^{j}}}}{dt}\frac{{\rm{D}}H}{\partial \overline{{{z}^{j}}}}}\] Actually, it implies a fact
\[\sum\limits_{j}{\frac{d{{z}^{j}}}{dt}\frac{d\overline{{{z}^{j}}}}{dt}}=\sum\limits_{j}{\frac{d\overline{{{z}^{j}}}}{dt}\frac{d{{z}^{j}}}{dt}}\]
According to corollary \ref{c4}, $H$ is a covariant conserved quantity.

By using \eqref{eq16} in terms of ${z}^{j}$, we obtain
\begin{align}
 \frac{\mathcal{D}}{dt}{{z}^{j}} &=\left\{ {{z}^{j}},H \right\}=2\sqrt{-1}\sum\limits_{j}{\left( -\frac{{\rm{D}}H}{\partial \overline{{{z}^{j}}}}+{{z}^{j}}\left( \frac{{\rm{D}}H}{\partial {{z}^{j}}}\frac{\partial s}{\partial \overline{{{z}^{j}}}}-\frac{{\rm{D}}H}{\partial \overline{{{z}^{j}}}}\frac{\partial s}{\partial {{z}^{j}}} \right) \right)} \notag\\
 & =-2\sqrt{-1}\frac{{\rm{D}}H}{\partial \overline{{{z}^{j}}}}+2\sqrt{-1}{{z}^{j}}\sum\limits_{j}{\left( \frac{{\rm{D}}H}{\partial {{z}^{j}}}\frac{\partial s}{\partial \overline{{{z}^{j}}}}-\frac{{\rm{D}}H}{\partial \overline{{{z}^{j}}}}\frac{\partial s}{\partial {{z}^{j}}} \right)}  \notag\\
 & =-2\sqrt{-1}\frac{{\rm{D}}H}{\partial \overline{{{z}^{j}}}}+{{z}^{j}}w  \notag
\end{align}
Here S-dynamics is represented by
\begin{equation}\label{eq14}
  w=2\sqrt{-1}\sum\limits_{j}{\left( \frac{{\rm{D}}H}{\partial {{z}^{j}}}\frac{\partial s}{\partial \overline{{{z}^{j}}}}-\frac{{\rm{D}}H}{\partial \overline{{{z}^{j}}}}\frac{\partial s}{\partial {{z}^{j}}} \right)}
\end{equation}
Similarly, the GCHS in terms of $\overline{{{z}^{j}}}$, it yields a result as theorem \ref{t8} given.
\begin{align}
 \frac{\mathcal{D}}{dt}\overline{{{z}^{j}}} & =\left\{ \overline{{{z}^{j}}},H \right\}=2\sqrt{-1}\sum\limits_{j}{\left( \frac{{\rm{D}}H}{\partial {{z}^{j}}}\frac{{\rm{D}}\overline{{{z}^{j}}}}{\partial \overline{{{z}^{j}}}}-\frac{{\rm{D}}\overline{{{z}^{j}}}}{\partial {{z}^{j}}}\frac{{\rm{D}}H}{\partial \overline{{{z}^{j}}}} \right)} \notag\\
 & =2\sqrt{-1}\sum\limits_{j}{\left( \frac{{\rm{D}}H}{\partial {{z}^{j}}}+\overline{{{z}^{j}}}\left( \frac{{\rm{D}}H}{\partial {{z}^{j}}}\frac{\partial s}{\partial \overline{{{z}^{j}}}}-\frac{\partial s}{\partial {{z}^{j}}}\frac{{\rm{D}}H}{\partial \overline{{{z}^{j}}}} \right) \right)} \notag\\
 & =2\sqrt{-1}\frac{{\rm{D}}H}{\partial {{z}^{j}}}+2\sqrt{-1}\overline{{{z}^{j}}}\sum\limits_{j}{\left( \frac{{\rm{D}}H}{\partial {{z}^{j}}}\frac{\partial s}{\partial \overline{{{z}^{j}}}}-\frac{\partial s}{\partial {{z}^{j}}}\frac{{\rm{D}}H}{\partial \overline{{{z}^{j}}}} \right)} \notag\\
 & =\frac{d}{dt}\overline{{{z}^{j}}}+\overline{{{z}^{j}}}w \notag
\end{align}
where TGHS follows $\frac{d}{dt}\overline{{{z}^{j}}}=2\sqrt{-1}\frac{{\rm{D}}H}{\partial {{z}^{j}}}$, while S-dynamics remains the same as \eqref{eq14} shown.

\subsection{S-dynamics and geometrio in complex coordinates}
In this section, we mainly discuss the relating properties of S-dynamics.  As previously defined and seen from theorem \ref{t1}, S-dynamics is completely expressed by $w={{\left\{ s,H \right\}}_{PB}}$ in which the Hamiltonian $H$ inside of the equation represents the total energy carried by matters or particles, it shows the matters themself, the function $s$ stands for the properties of the structure of the space or the manifolds as a stage of motion for matters. Just as Eq \eqref{eq4} illustrated, it has
\[\frac{\mathcal{D}}{dt}1=w={{\left\{ s,H \right\}}_{PB}}=2\sqrt{-1}\sum\limits_{j}{\left( \frac{\partial s}{\partial \overline{{{z}^{j}}}}\frac{\partial H}{\partial {{z}^{j}}}-\frac{\partial s}{\partial {{z}^{j}}}\frac{\partial H}{\partial \overline{{{z}^{j}}}} \right)}\]where $\frac{\partial s}{\partial {{z}^{j}}},~~\frac{\partial s}{\partial \overline{{{z}^{j}}}}$ are the partial derivative with respect to the coordinates $q$ and momentum $p$.

Actually, let's consider
\[{{\left\{ s,f \right\}}_{PB}}=2\sqrt{-1}\sum\limits_{j}{\left( \frac{\partial s}{\partial \overline{{{z}^{j}}}}\frac{\partial f}{\partial {{z}^{j}}}-\frac{\partial s}{\partial {{z}^{j}}}\frac{\partial f}{\partial \overline{{{z}^{j}}}} \right)}\]for function $f$, then let $f$ be taken as
${{z}^{j}}, ~\overline{{{z}^{j}}} $ respectively, we obtain
\begin{align}\label{eq15}
  & {{\left\{ s,{{z}^{j}} \right\}}_{PB}}=2\sqrt{-1}{\frac{\partial s}{\partial \overline{{{z}^{j}}}}} \\
 & {{\left\{ s,\overline{{{z}^{j}}} \right\}}_{PB}}=-2\sqrt{-1}{\frac{\partial s}{\partial {{z}^{j}}}} \notag
\end{align}
The geometrio shown by \ref{t4} in complex coordinates is given by
\eqref{eq15},  and $w={{\left\{ s,H \right\}}_{PB}}$.

\subsection{The connection between the TGHS and the S-dynamics}
Recall \eqref{eq16}, the covariant time operator reads $$\frac{\mathcal{D}}{dt}=\left\{\cdot ,H \right\}=\sum\limits_{j}{\left( \frac{d\overline{{{z}^{j}}}}{dt}\frac{{\rm{D}}}{\partial \overline{{{z}^{j}}}}+\frac{d{{z}^{j}}}{dt}\frac{{\rm{D}}}{\partial {{z}^{j}}} \right)}$$Furthermore, it further yields
\begin{align}
 \frac{\mathcal{D}}{dt} & =\left\{\cdot ,H \right\}=2\sqrt{-1}\sum\limits_{j}{\left( \frac{{\rm{D}}H}{\partial {{z}^{j}}}\frac{\partial }{\partial \overline{{{z}^{j}}}}-\frac{{\rm{D}}H}{\partial \overline{{{z}^{j}}}}\frac{\partial }{\partial {{z}^{j}}}+\frac{{\rm{D}}H}{\partial {{z}^{j}}}\frac{\partial s}{\partial \overline{{{z}^{j}}}}-\frac{{\rm{D}}H}{\partial \overline{{{z}^{j}}}}\frac{\partial s}{\partial {{z}^{j}}} \right)}\notag \\
 & =2\sqrt{-1}\sum\limits_{j}{\left( \frac{{\rm{D}}H}{\partial {{z}^{j}}}\frac{\partial }{\partial \overline{{{z}^{j}}}}-\frac{{\rm{D}}H}{\partial \overline{{{z}^{j}}}}\frac{\partial }{\partial {{z}^{j}}} \right)}+2\sqrt{-1}\sum\limits_{j}{\left( \frac{{\rm{D}}H}{\partial {{z}^{j}}}\frac{\partial s}{\partial \overline{{{z}^{j}}}}-\frac{{\rm{D}}H}{\partial \overline{{{z}^{j}}}}\frac{\partial s}{\partial {{z}^{j}}} \right)} \notag\\
 & =\frac{d}{dt}+w \notag
\end{align}
where the thorough time operator is precisely expressed as
\begin{align}
  \frac{d}{dt}&=2\sqrt{-1}\sum\limits_{j}{\left( \frac{{\rm{D}}H}{\partial {{z}^{j}}}\frac{\partial }{\partial \overline{{{z}^{j}}}}-\frac{{\rm{D}}H}{\partial \overline{{{z}^{j}}}}\frac{\partial }{\partial {{z}^{j}}} \right)} \notag\\
 & =2\sqrt{-1}\sum\limits_{j}{\left( \frac{\partial H}{\partial {{z}^{j}}}\frac{\partial }{\partial \overline{{{z}^{j}}}}-\frac{\partial H}{\partial \overline{{{z}^{j}}}}\frac{\partial }{\partial {{z}^{j}}} \right)} \notag\\
 & \begin{matrix}
   {} & {} & {} & {}  \\
\end{matrix}+2\sqrt{-1}H\sum\limits_{j}{\left( \frac{\partial s}{\partial {{z}^{j}}}\frac{\partial }{\partial \overline{{{z}^{j}}}}-\frac{\partial s}{\partial \overline{{{z}^{j}}}}\frac{\partial }{\partial {{z}^{j}}} \right)}\notag
\end{align}
and in the form expressed by
\[\frac{d}{dt}=\sum\limits_{j}{\left( \frac{d\overline{{{z}^{j}}}}{dt}\frac{\partial }{\partial \overline{{{z}^{j}}}}+\frac{d{{z}^{j}}}{dt}\frac{\partial }{\partial {{z}^{j}}} \right)}\]
Therefore, for any given function $f$, the TGHS about it is given by
\begin{align}
  \frac{d}{dt}f& =2\sqrt{-1}\sum\limits_{j}{\left( \frac{{\rm{D}}H}{\partial {{z}^{j}}}\frac{\partial f}{\partial \overline{{{z}^{j}}}}-\frac{{\rm{D}}H}{\partial \overline{{{z}^{j}}}}\frac{\partial f}{\partial {{z}^{j}}} \right)} \notag\\
 & =\sum\limits_{j}{\left( \frac{d\overline{{{z}^{j}}}}{dt}\frac{\partial f}{\partial \overline{{{z}^{j}}}}+\frac{d{{z}^{j}}}{dt}\frac{\partial f}{\partial {{z}^{j}}} \right)} \notag
\end{align}
Hence, the relation between the TGHS and the S-dynamics is given by
\[w=\sum\limits_{j}{\left( \frac{d\overline{{{z}^{j}}}}{dt}\frac{\partial s}{\partial \overline{{{z}^{j}}}}+\frac{d{{z}^{j}}}{dt}\frac{\partial s}{\partial {{z}^{j}}} \right)}=\sum\limits_{j}{\left( \begin{matrix}
   \frac{d{{z}^{j}}}{dt} & \frac{d\overline{{{z}^{j}}}}{dt}  \\
\end{matrix} \right)\left( \begin{matrix}
   \frac{\partial s}{\partial {{z}^{j}}}  \\
   \frac{\partial s}{\partial \overline{{{z}^{j}}}}  \\
\end{matrix} \right)}\]it's another equivalent expression of \eqref{eq14}.

\begin{theorem}\label{t7}
The GCHS, TGHS and S-dynamics:
\begin{description}
  \item[GCHS:]
  $\frac{\mathcal{D}f}{dt}=\left\{f ,H \right\}=\sum\limits_{j}{\left( \frac{d\overline{{{z}^{j}}}}{dt}\frac{{\rm{D}}f}{\partial \overline{{{z}^{j}}}}+\frac{d{{z}^{j}}}{dt}\frac{{\rm{D}}f}{\partial {{z}^{j}}} \right)}$.
  \item[TGHS:]
$\frac{d}{dt}f=\sum\limits_{j}{\left( \frac{d\overline{{{z}^{j}}}}{dt}\frac{\partial f}{\partial \overline{{{z}^{j}}}}+\frac{d{{z}^{j}}}{dt}\frac{\partial f}{\partial {{z}^{j}}} \right)}$.
  \item[S-dynamics:]
$w=\sum\limits_{j}{\left( \frac{d\overline{{{z}^{j}}}}{dt}\frac{\partial s}{\partial \overline{{{z}^{j}}}}+\frac{d{{z}^{j}}}{dt}\frac{\partial s}{\partial {{z}^{j}}} \right)}$.
\end{description}
where TGHS in terms of ${z}^{j}$ and $\overline{{{z}^{j}}}$ given by $\frac{d{{z}^{j}}}{dt}, \frac{d}{dt}\overline{{{z}^{j}}}$ are by using GSPB in complex coordinates in the form as theorems \ref{t1} and \ref{t8} shown respectively.
\end{theorem}
According to corollary \ref{c4}, the covariant equilibrium equation of the GCHS in complex coordinates in terms of a given function $f$ can be rewritten in the form
\[\sum\limits_{j}{\frac{d\overline{{{z}^{j}}}}{dt}\frac{{\rm{D}}f}{\partial \overline{{{z}^{j}}}}}+\sum\limits_{j}{\frac{d{{z}^{j}}}{dt}\frac{{\rm{D}}f}{\partial {{z}^{j}}}}=0\]That is,
$\sum\limits_{j}{\frac{d\overline{{{z}^{j}}}}{dt}\frac{{\rm{D}}f}{\partial \overline{{{z}^{j}}}}}=-\sum\limits_{j}{\frac{d{{z}^{j}}}{dt}\frac{{\rm{D}}f}{\partial {{z}^{j}}}}$.  Similarly, for the equilibrium equation of the TGHS, it gives rise to the result,
\[\sum\limits_{j}{\left( \frac{d\overline{{{z}^{j}}}}{dt}\frac{\partial f}{\partial \overline{{{z}^{j}}}}+\frac{d{{z}^{j}}}{dt}\frac{\partial f}{\partial {{z}^{j}}} \right)}=0\]or the equality
$\sum\limits_{j}{\frac{d\overline{{{z}^{j}}}}{dt}\frac{\partial f}{\partial \overline{{{z}^{j}}}}}=-\sum\limits_{j}{\frac{d{{z}^{j}}}{dt}\frac{\partial f}{\partial {{z}^{j}}}}$. Meanwhile, S-dynamics $w\neq 0$ always holds.

\subsection{Acceleration-like in complex coordinates}

\begin{theorem}
  Acceleration-like in complex coordinates is given by
\[\frac{{{\mathcal{D}}^{2}}}{d{{t}^{2}}}f=\frac{{{d}^{2}}}{d{{t}^{2}}}f+2w\frac{d}{dt}f+
f\beta\]for the function $f$.
  where
$\frac{d}{dt}f=\sum\limits_{j}{\left( \frac{d\overline{{{z}^{j}}}}{dt}\frac{\partial f}{\partial \overline{{{z}^{j}}}}+\frac{d{{z}^{j}}}{dt}\frac{\partial f}{\partial {{z}^{j}}} \right)}$ is the TGHS, and $w$ is the S-dynamics, $\beta =\frac{d}{dt}w+{{w}^{2}}$.
\begin{proof}
  Since the GCHS in complex coordinates is described by theorem \ref{t2}, then
\begin{align}
 \frac{{{\mathcal{D}}^{2}}}{d{{t}^{2}}}f &=\left\{ \frac{\mathcal{D}}{dt}f,H \right\}={{\left\{ \frac{\mathcal{D}}{dt}f,H \right\}}_{PB}}+G\left( s,\frac{\mathcal{D}}{dt}f,H \right) \notag\\
 & =2\sqrt{-1}\sum\limits_{j}{\left( \frac{{\rm{D}}H}{\partial {{z}^{j}}}\frac{{\rm{D}}}{\partial \overline{{{z}^{j}}}}\frac{\mathcal{D}}{dt}f-\frac{{\rm{D}}H}{\partial \overline{{{z}^{j}}}}\frac{{\rm{D}}}{\partial {{z}^{j}}}\frac{\mathcal{D}}{dt}f \right)} \notag
\end{align}
where geometric bracket in terms of $f,H$ reads
  \[G\left( s,\frac{\mathcal{D}}{dt}f,H \right)=\frac{\mathcal{D}f}{dt}w+H{{\left\{ \frac{\mathcal{D}}{dt}f,s \right\}}_{PB}}\]
Actually, acceleration-like is given by \[\frac{{{\mathcal{D}}^{2}}}{d{{t}^{2}}}f=\frac{{{d}^{2}}}{d{{t}^{2}}}f+2w\frac{d}{dt}f+f\left( \frac{d}{dt}w+{{w}^{2}} \right)\]
 or by using theorem \ref{t7}, we can show acceleration-like
\begin{align}
\frac{{{\mathcal{D}}^{2}}}{d{{t}^{2}}}f  & =\sum\limits_{j}{\left( \frac{d\overline{{{z}^{j}}}}{dt}\frac{{\rm{D}}}{\partial \overline{{{z}^{j}}}}\frac{\mathcal{D}}{dt}f+\frac{d{{z}^{j}}}{dt}\frac{{\rm{D}}}{\partial {{z}^{j}}}\frac{\mathcal{D}}{dt}f \right)}  \notag\\
 & =\sum\limits_{j}{\left( \frac{d\overline{{{z}^{j}}}}{dt}\frac{{\rm{D}}}{\partial \overline{{{z}^{j}}}}\frac{d}{dt}f+\frac{d{{z}^{j}}}{dt}\frac{{\rm{D}}}{\partial {{z}^{j}}}\frac{d}{dt}f+\frac{d\overline{{{z}^{j}}}}{dt}\frac{{\rm{D}}}{\partial \overline{{{z}^{j}}}}\left( fw \right)+\frac{d{{z}^{j}}}{dt}\frac{{\rm{D}}}{\partial {{z}^{j}}}\left( fw \right) \right)}  \notag\\
 & =\sum\limits_{j}{\left( \frac{d\overline{{{z}^{j}}}}{dt}\frac{{\rm{D}}}{\partial \overline{{{z}^{j}}}}\frac{d}{dt}f+\frac{d{{z}^{j}}}{dt}\frac{{\rm{D}}}{\partial {{z}^{j}}}\frac{d}{dt}f \right)}+\sum\limits_{j}{\left( \frac{d\overline{{{z}^{j}}}}{dt}\frac{{\rm{D}}}{\partial \overline{{{z}^{j}}}}\left( fw \right)+\frac{d{{z}^{j}}}{dt}\frac{{\rm{D}}}{\partial {{z}^{j}}}\left( fw \right) \right)}  \notag\\
 & =\frac{\mathcal{D}}{dt}\frac{d}{dt}f+\frac{\mathcal{D}}{dt}\left( fw \right)  \notag
\end{align}

\end{proof}

\end{theorem}

According to corollary \ref{c4}, we can immediately deduce a conclusion as follows:
\begin{corollary}\label{c5}
The $\frac{\mathcal{D}}{dt}f=\left\{f ,H \right\}$ is a covariant conserved quantity if and only if $\frac{{{\mathcal{D}}^{2}}}{d{{t}^{2}}}f=0$.
\end{corollary}
Above corollary means \[\frac{{{d}^{2}}}{d{{t}^{2}}}f+2w\frac{d}{dt}f+
f\beta=0\]holds for covariant conserved quantity $\frac{\mathcal{D}}{dt}f=\left\{f ,H \right\}$.

Correspondingly, the acceleration-like in terms of ${{z}^{j}}$ is
\[\frac{{{\mathcal{D}}^{2}}}{d{{t}^{2}}}{{z}^{j}}=\frac{{{d}^{2}}}{d{{t}^{2}}}{{z}^{j}}+2w\frac{d}{dt}{{z}^{j}}+{{z}^{j}}\beta \]In accordance with corollary \ref{c5}, $\frac{\mathcal{D}}{dt}{{z}^{j}}$ is a covariant conserved quantity, then the equation for it is
$$\frac{{{d}^{2}}}{d{{t}^{2}}}{{z}^{j}}+2w\frac{d}{dt}{{z}^{j}}+{{z}^{j}}\beta =0 $$
\begin{remark}
The covariant time operator and thorough time operator are listed as follows:
  \begin{description}
  \item[Covariant time operator:]
  $\frac{\mathcal{D}}{dt}=\sum\limits_{j}{\left( \frac{d\overline{{{z}^{j}}}}{dt}\frac{{\rm{D}}}{\partial \overline{{{z}^{j}}}}+\frac{d{{z}^{j}}}{dt}\frac{{\rm{D}}}{\partial {{z}^{j}}} \right)}=\sum\limits_{j}{\left( \frac{d{{z}^{j}}}{dt},\frac{d\overline{{{z}^{j}}}}{dt} \right){{\left( \begin{matrix}
   \frac{\rm{D} }{\partial {{z}^{j}}} & \frac{\rm{D} }{\partial \overline{{{z}^{j}}}}  \\
\end{matrix} \right)}^{T}}}$.
  \item[Thorough time operator:]
$\frac{d}{dt}=\sum\limits_{j}{\left( \frac{d\overline{{{z}^{j}}}}{dt}\frac{\partial }{\partial \overline{{{z}^{j}}}}+\frac{d{{z}^{j}}}{dt}\frac{\partial }{\partial {{z}^{j}}} \right)}=\sum\limits_{j}{\left( \frac{d{{z}^{j}}}{dt},\frac{d\overline{{{z}^{j}}}}{dt} \right){{\left( \begin{matrix}
   \frac{\partial }{\partial {{z}^{j}}} & \frac{\partial }{\partial \overline{{{z}^{j}}}}  \\
\end{matrix} \right)}^{T}}}$.
\end{description}
where TGHS in terms of ${z}^{j}$ and $\overline{{{z}^{j}}}$ are given by $\frac{d{{z}^{j}}}{dt}, \frac{d}{dt}\overline{{{z}^{j}}}$, respectively.
\end{remark}

\section{Conclusions}
In this paper, we have proposed new formula to further develop the classical HS to extend it much more covariant as GCHS shown in real coordinates.  An attempt to rebuild the framework of known theory of Hamiltonian system, and it obviously works, the GSPB and GCHS that have been proposed which represent and depict a whole Hamiltonian picture one has ever faced in the real situation. A complete Hamiltonian system can be completely encapsulated by GCHS and its subsystem.

By introducing the complex structural function $s$ only associated with domain or structure of manifold, one has expanded the Hamiltonian system to the general covariant form which can be applied to general non-Euclidean case.  The GSPB which has generalized the PB is compatible with the GCHS that contains two dynamical subsystem: TGHS and S-dynamics which are independent system to describe the corresponding dynamical properties, but for the entire system, they are together contributed their parts.  As a result, one can easily see that GCHS with GSPB in complex coordinates: TGHS and S-dynamics all keep the mathematical form invariant. One has come up with an idea, a new theory to remedy the classical theory, it contrives an advanced framework for a complete theory. S-dynamics is devised simultaneously which reflects deeply on a Hamiltonian subject.


\begin{thebibliography}{99}
\bibitem{1}Pauli  W. On the Hamiltonian structure of non-local field theories [J]. IL Nuovo Cimento.  1953, 10: 648-667.

\bibitem{2}
J D. Meiss. Differential Dynamical Systems [M]. Mathematical Modeling and Computation, 2007.


\bibitem{3}Stephen W , Mazel D S. Introduction to Applied Nonlinear Dynamical Systems and Chaos [J]. Computers in Physics. 1990, 4(5): 563-580.


\bibitem{4}Weinstein. A. The Local Structure of Possion Manifold [J]. Diff. Geom. 1983, 18: 523-557.

\bibitem{5}
 Wang G. A study of generalized covariant Hamilton systems on generalized Poisson manifold [J].
\href{https://arxiv.org/abs/1710.10597}{arXiv:1710.10597v7}
 \bibitem{6}
Strocchi, F. Rev. Mod. Phys. 1966 (38), 36.

\bibitem{7}
Wang G. Analogy between geodesic equation and the GCHS on Riemannian manifolds [J].
\href{https://arxiv.org/abs/2002.10825}{arXiv:2002.10825v1}



\end{thebibliography}
\end{document}